\documentclass[12pt]{iopart}

\usepackage{graphicx} 
\begin{document}

\title[Improvement Studies of an Effective Interaction]{Improvement Studies of an Effective Interaction for N=Z sd-shell Nuclei by Neural Networks}

\author{Serkan Akkoyun$^1$ $^2$, Nadjet Laouet$^3$, Fatima Benrachi$^3$}

\address{$^1$ Department of Physics, Faculty of Sciences, Sivas Cumhuriyet University, Sivas, Turkey}
\address{$^2$ Artificial Intelligence Systems and Data Science Application and Research Center, Sivas Cumhuriyet University, Sivas, Turkey}
\address{$^3$ LPMPS Laboratory, Frères Mentouri Constantine-1 University, 25017 Constantine, Algeria}

\ead{sakkoyun@cumhuriyet.edu.tr}
\vspace{10pt}
\begin{indented}
\item[]January 2020
\end{indented}

\begin{abstract}
The nuclear shell model is one of the successful models in theoretical understanding of nuclear structure. If a convenient effective interaction can be found between nucleons, various observables such as energies of nuclear states are accurately predicted by this model. The basic requirements for the shell model calculations are a set of single particle energies and two-body interaction matrix elements (TBME) which construct the residual interaction between nucleons. This latter could be parameterized in different ways. In this study, we have used a different approach to improve existing USD type Hamiltonians for the shell model calculations of $N=Z$ nuclei in the A=16-40 region. After obtaining the SDNN new effective interaction, shell model calculations have been performed for all $N=Z$ nuclei in sd shell. In which, $^{16}{O}$ doubly magic nucleus has been assumed as an inert core and active particles are distributed in the $d_{5/2}$, $s_{1/2}$ and $d_{3/2}$ single particle orbits. The rms deviations from experimental energy values are lower for the newly generated effective interaction than those obtained using the original one for the studied nuclei. 
\end{abstract}

%
\vspace{0.1pc}

%
\submitto{\JPG}
%
%
%

\section{Introduction}

The nuclear shell model is a successful tool in nuclear structure comprehension. The sd shell nuclei have received intense focus in nuclear physics studies. Nuclear excited energy levels are confidently calculated by means of this model. Similar to the atomic shell model, nuclear properties can be described using valence nucleons situated out of closed shells (with magic numbers 2, 8, 20, 28, 50, 82 and 126). In nuclear shell model, nucleons are considered to be moved in an average potential and interact with each other through a residual or effective interaction. This interaction between the nucleons has mainly two-body character. Under the assumptions where the total angular momentum (\textit{J}) is a good quantum number, many calculations have been done on the sd-shell space model in order to get systematic explanation of the low-lying levels in the nuclei from oxygen to calcium \cite{Talmi}. The main objective is to obtain a set of effective interaction two body matrix elements (TBME) between nucleons. These elements might be useful to obtain better nuclear force understanding and to predict unknown nuclear quantities.

One of the commonly used Hamiltonian in sd space model is USD Hamiltonian \cite{Brown1}. This latter defined by 63 TBMEs, has provided realistic sd-shell wave functions for the usage in nuclear structure models, nuclear spectroscopy and nuclear astrophysics \cite{Brown1}. Their values are derived from the experimental binding energies and excitation energies for nuclei in the region A = 16–40. The original USD Hamiltonian \cite{Wildenthal, Brown2} was obtained from a least squares fit of 380 energy data from 66 nuclei. In the improved USD Hamiltonians (USDA and USDB), 608 states in 77 nuclei distributed over sd-shell are considered. The new Hamiltonians are more sensitive for realistic shell-model wave functions.

In recent years, Artificial Neural Network (ANN) \cite{Haykin} has been used in many fields in nuclear physics. It has been used successfully for developing nuclear mass systematic \cite{Athanassopoulos, Bayram}, determination of proton radius \cite{Graczyk1}, nuclear mass prediction \cite{Utama1}, obtaining fission barrier heights \cite{Akkoyun1}, obtaining nuclear charge radii \cite{Akkoyun2, Utama2} and predictions of hadronic model \cite{Graczyk2}. Since this method is successful in understanding non-linear relationship between input and output data, layered feed-forward ANN can be used to generate TBME for the calculations for $N=Z$ nuclei in sd shell. 

In this paper, USDB interaction Hamiltonian was re-estimated by using artificial neural networks. After generation a new effective interaction SDNN, both original and newly obtained Hamiltonians were used to calculate the energy levels of $^{18}{F}$, $^{20}{Ne}$, $^{22}{Na}$, $^{24}{Mg}$, $^{26}{Al}$, $^{28}{Si}$, $^{30}{P}$, $^{32}{S}$, $^{34}{Cl}$, $^{36}{Ar}$ and $^{38}{K}$ isotopes by using KShell nuclear structure shell model code \cite{Shimizu}. According to the calculations, the obtained results from SDNN are closer to the experimental values than USDB for $N=Z$ nuclei in sd region. In section 2, the brief summaries for artificial neural networks and shell model calculations will be given. In section 3, the results for different Hamiltonians are discussed. Finally, Section 4 presents our conclusion. Finally, Section 4 presents our conclusions.

\section{Material and Methods}

\subsection{Artificial Neural Networks}

Artificial neural networks (ANN) is one of artificial intelligence sub area \cite{Haykin}. ANN mimics the human brain functionality in order to give outputs as consequence of the computation of the inputs. It is composed of processing units called neurons which have adaptive synaptic weights. The main task is to determine appropriate weights belonging to the problem under consideration. ANN consists of three layers: input, hidden and output. The number of hidden layers can differ, but a single hidden layer is generally enough for almost all problems \cite{Hornik}.
 
In this work, ANN (1-8-1) architecture composed with one input layer with one neuron, one hidden layer with 8 neurons and one output layer with one neuron, was used. From 1 to 10 hidden neuron numbers, all the architectures have been used to generate new interaction data set. When the number of hidden neuron is greater than 10, ANN starts to memorize the data. The best results were obtained for 8 neurons in hidden layer. The input neurons correspond to USDB TBMEs besides output neurons are new matrix elements (SDNN). The total numbers of adjustable weight was 16 according to the formula $p.h+h.r=2h$. Here $p$, $h$ and $r$ are the input, hidden and output neuron numbers, respectively. . The input neuron collects problem data from environment and transmits them via weighted connections to the hidden neurons and then to the output ones. In the calculations, the hidden neuron activation function was chosen as hyperbolic tangent for hidden layer.

ANN consists of two main stages. One of them is training stage and the other is test one. Back-propagation algorithm with Levenberg-Marquardt \cite{Levenberg, Marquardt} for the training of the ANN was used. ANN modifies its weights until an acceptable error level between predicted and desired outputs. The error function which measures the difference between outputs was mean square error. After the training is completed, the performance of the network is tested over data by final weights. If the mean square error of the test data is low enough, the ANN is considered to have learned the functional relationship between input and output data \cite{Yildiz}. In this study, the mean square error value was about 6x10-29. 

\subsection{Shell Model Calculations}
For the shell model calculations by using both USDB and newly generated SDNN matrix elements, KShell nuclear structure shell model code \cite{Shimizu} has been used under M-scheme representation with the thick-restart Lanczos method. The code is a powerful tool in order to calculate energy levels, spin, isospin, magnetic and quadrupole moments, E2/M1 transition probabilities, and one-particle spectroscopic factors in the nuclei. The code is able to carry out calculations for dimension up to tens of billions if enough memory is available on the computers. Besides this code, there are several codes written for the shell model calculations are existing such as NuShellX \cite{Nushell}, Redstick \cite{Redstick}, Bigstick \cite{Bigstick}, Antoine \cite{Antoine}, Oxbash \cite{Oxbash}, etc.

In this study, the excited energy levels of $^{18}{F}$, $^{20}{Ne}$, $^{22}{Na}$, $^{24}{Mg}$, $^{26}{Al}$, $^{28}{Si}$, $^{30}{P}$, $^{32}{S}$, $^{34}{Cl}$, $^{36}{Ar}$ and $^{38}{K}$ isotopes in the sd Single Particle Space (SPS) model have been calculated by using KShell code. This SPS consists of $d_{5/2}$, $s_{1/2}$ and $d_{3/2}$ valence orbitals above the doubly magic $^{16}{O}$ core.
 
Valence nucleons move in model \textit{j}-orbit space and their Hamiltonian is given by 

\vspace{1pc}
$H=E_{0}+\Sigma \varepsilon_{i} a_{i}^{+} a_{j} + (1/2) \Sigma \langle ij \vert V \vert kl \rangle a_{i}^{+} a_{j}^{+} a_{k} a_{l}$
\vspace{1pc}
\newline
where $E_{0}$ denotes the inert core binging energy, $\varepsilon_{i}$  is the valence orbit single particle energies (SPEs and $\langle ij \vert V \vert kl \rangle$ term is the residual interaction -between the valence particles- two-body matrix elements (TBMEs). Here \textit{i}, \textit{j}, \textit{k}, \textit{l} are single particle states, $a_{i}$ and $a_{i}^{+}$ are annihilation and creation operators, respectively. The taken SPEs are $+2.1117$, $-3.9257$ and $-3.2079$ for $d_{5/2}$, $s_{1/2}$ and $d_{3/2}$ orbitals, respectively.

\section{Results and Discussion}
In this work, USDB interaction TBMEs are used as the ANN inputs. By using ANN method, we have generated new set of SDNN interaction TBMEs. As shown in Table 1 and Table 2 for $T=0$ and $T=1$, respectively, the new matrix elements are slightly different from original ones. The root mean square (rms) deviation from original matrix elements is $0.052$ MeV. For $\langle 11 \vert V \vert 33 \rangle_{01}$ matrix element, the difference from original matrix elements is $0$, whereas the value reaches the maximum as $-0.2768$ for $\langle 21 \vert V \vert 21 \rangle_{21}$ matrix element.

\begin{table}
\caption{\label{Table1}USDB and newly generated SDNN effective interaction matrix elements $v(i,j,k,l,J,T)$ in MeV for $T=0$. The orbits are labeled by 1=$d_{5/2}$, 2=$s_{1/2}$ and 3=$d_{3/2}$.}
\centering
\footnotesize
\begin{tabular}{@{}llllllll}
\br
i&j&k&l&J&T&USDB&SDNN\\
\mr
2	&	2	&	2	&	2	&	1	&	0	&	-1.3796	&	-1.3781	\\
2	&	2	&	2	&	1	&	1	&	0	&	3.4987	&	3.4872	\\
2	&	2	&	1	&	1	&	1	&	0	&	1.6647	&	1.6639	\\
2	&	2	&	1	&	3	&	1	&	0	&	0.0272	&	0.0209	\\
2	&	2	&	3	&	3	&	1	&	0	&	-0.5344	&	-0.5367	\\
2	&	1	&	2	&	1	&	1	&	0	&	-6.0099	&	-6.0196	\\
2	&	1	&	1	&	1	&	1	&	0	&	0.1922	&	0.1355	\\
2	&	1	&	1	&	3	&	1	&	0	&	1.6231	&	1.6454	\\
2	&	1	&	3	&	3	&	1	&	0	&	2.0226	&	2.0220	\\
1	&	1	&	1	&	1	&	1	&	0	&	-1.6582	&	-1.6629	\\
1	&	1	&	1	&	3	&	1	&	0	&	-0.8493	&	-0.8570	\\
1	&	1	&	3	&	3	&	1	&	0	&	0.1574	&	0.1589	\\
1	&	3	&	1	&	3	&	1	&	0	&	-4.0460	&	-3.9766	\\
1	&	3	&	3	&	3	&	1	&	0	&	-0.9201	&	-0.9159	\\
3	&	3	&	3	&	3	&	1	&	0	&	-3.7093	&	-3.7230	\\
2	&	1	&	2	&	1	&	2	&	0	&	-4.2117	&	-4.2126	\\
2	&	1	&	2	&	3	&	2	&	0	&	-0.6464	&	-0.6348	\\
2	&	1	&	1	&	3	&	2	&	0	&	-0.4429	&	-0.4403	\\
2	&	3	&	2	&	3	&	2	&	0	&	-0.3154	&	-0.2893	\\
2	&	3	&	1	&	3	&	2	&	0	&	-2.5110	&	-2.5453	\\
1	&	3	&	1	&	3	&	2	&	0	&	-1.8504	&	-1.8958	\\
2	&	2	&	2	&	2	&	3	&	0	&	-1.6651	&	-1.6653	\\
2	&	2	&	2	&	1	&	3	&	0	&	2.3102	&	2.2873	\\
2	&	2	&	2	&	3	&	3	&	0	&	-1.2167	&	-1.2114	\\
2	&	2	&	1	&	1	&	3	&	0	&	1.1792	&	1.1861	\\
2	&	1	&	2	&	1	&	3	&	0	&	-1.2124	&	-1.2180	\\
2	&	1	&	2	&	3	&	3	&	0	&	1.2526	&	1.2300	\\
2	&	1	&	1	&	1	&	3	&	0	&	1.4300	&	1.4413	\\
2	&	3	&	2	&	3	&	3	&	0	&	-4.1823	&	-4.1781	\\
2	&	3	&	1	&	1	&	3	&	0	&	0.0968	&	0.2010	\\
1	&	1	&	1	&	1	&	3	&	0	&	-2.9660	&	-2.9538	\\
2	&	1	&	2	&	1	&	4	&	0	&	-4.6189	&	-4.6085	\\
2	&	2	&	2	&	2	&	5	&	0	&	-4.3205	&	-4.3196	\\
\br
\end{tabular}\\
\end{table}

\begin{table}
\caption{\label{Table1}USDB and newly generated SDNN effective interaction matrix elements $v(i,j,k,l,J,T)$ in MeV for $T=1$. The orbits are labeled by 1=$d_{5/2}$, 2=$s_{1/2}$ and 3=$d_{3/2}$.}
\centering
\footnotesize
\begin{tabular}{@{}llllllll}
\br
i&j&k&l&J&T&USDB&SDNN\\
\mr
2	&	2	&	2	&	2	&	0	&	1	&	-2.5598	&	-2.6188	\\
2	&	2	&	1	&	1	&	0	&	1	&	-3.1025	&	-3.1129	\\
2	&	2	&	3	&	3	&	0	&	1	&	-1.5602	&	-1.6112	\\
1	&	1	&	1	&	1	&	0	&	1	&	-1.8992	&	-1.9019	\\
1	&	1	&	3	&	3	&	0	&	1	&	-1.0150	&	-1.0150	\\
3	&	3	&	3	&	3	&	0	&	1	&	-1.6913	&	-1.6966	\\
2	&	1	&	2	&	1	&	1	&	1	&	0.6556	&	0.6362	\\
2	&	1	&	1	&	3	&	1	&	1	&	-0.0456	&	-0.0599	\\
1	&	3	&	1	&	3	&	1	&	1	&	0.5158	&	0.5349	\\
2	&	2	&	2	&	2	&	2	&	1	&	-1.0007	&	-0.9459	\\
2	&	2	&	2	&	1	&	2	&	1	&	-0.2137	&	-0.2430	\\
2	&	2	&	2	&	3	&	2	&	1	&	-0.9317	&	-0.8752	\\
2	&	2	&	1	&	1	&	2	&	1	&	-1.2187	&	-1.2095	\\
2	&	2	&	1	&	3	&	2	&	1	&	0.8866	&	0.8886	\\
2	&	1	&	2	&	1	&	2	&	1	&	-0.1545	&	0.1223	\\
2	&	1	&	2	&	3	&	2	&	1	&	-0.3147	&	-0.3063	\\
2	&	1	&	1	&	1	&	2	&	1	&	-0.5032	&	-0.5210	\\
2	&	1	&	1	&	3	&	2	&	1	&	0.3713	&	0.3776	\\
2	&	3	&	2	&	3	&	2	&	1	&	-0.9405	&	-0.9211	\\
2	&	3	&	1	&	1	&	2	&	1	&	-0.3173	&	-0.3817	\\
2	&	3	&	1	&	3	&	2	&	1	&	1.6131	&	1.6118	\\
1	&	1	&	1	&	1	&	2	&	1	&	-0.0974	&	-0.0988	\\
1	&	1	&	1	&	3	&	2	&	1	&	0.3494	&	0.3475	\\
1	&	3	&	1	&	3	&	2	&	1	&	-0.3034	&	-0.3286	\\
2	&	1	&	2	&	1	&	3	&	1	&	0.7673	&	0.5537	\\
2	&	1	&	2	&	3	&	3	&	1	&	-0.5525	&	-0.5559	\\
2	&	3	&	2	&	3	&	3	&	1	&	0.6841	&	0.6663	\\
2	&	2	&	2	&	2	&	4	&	1	&	-0.2069	&	-0.2135	\\
2	&	2	&	2	&	1	&	4	&	1	&	-1.3349	&	-1.3440	\\
2	&	1	&	2	&	1	&	4	&	1	&	-1.4447	&	-1.4464	\\
\br
\end{tabular}\\
\end{table}

\normalsize

\begin{figure}
\centering
\includegraphics[height=23cm]{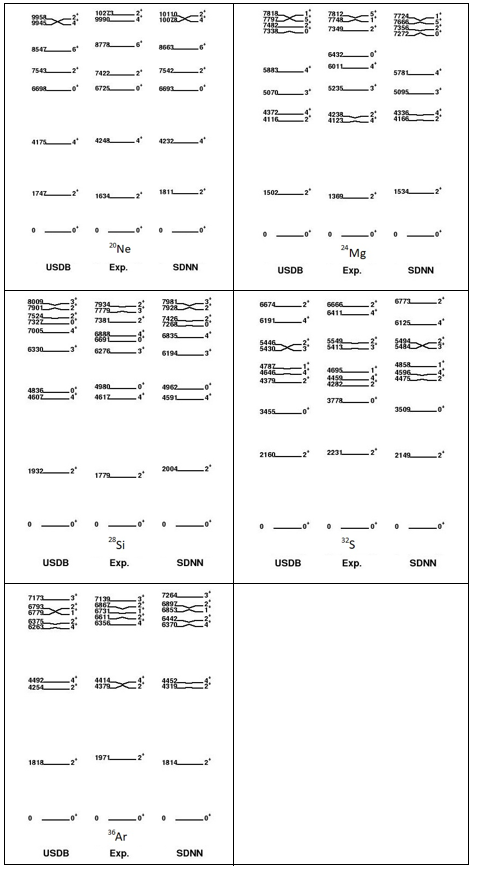}
\caption{Experimental and calculated energies for even-even $N=Z$ sd-shell nuclei by using USDB and SDNN interactions.}
\label{fig1}
\end{figure}

In the beginning of this study, we have calculated the first ten positive parity energy levels and spins of even-even $N=Z$ nuclei in sd-shell by using both USDB and SDNN interactions. All the ground state spin/parity values were reproduced correctly by using the new SDNN interaction. As shown in Fig.\ref{fig1}, the calculated energy level deviations from the experimental values \cite{nudat} are generally smaller for SDNN interaction. For $^{20}{Ne}$, $^{24}{Mg}$, $^{28}{Si}$, $^{32}{S}$ and $^{36}{Ar}$ isotopes, root mean square (rms) deviation values from the experimental energy levels are respectively $163.917$, $330.292$, $244.900$, $156.905$ and $121.655$ MeV from the calculations using USDB interaction. Whereas these values from the SDNN interaction are $116.221$, $312.307$, $220.341$, $170.940$ and $105.983$ MeV. Only for $^{32}{S}$ isotope, USDB interaction gives better results than SDNN. The rms deviations for the all even-even $N=Z$ isotopes in sd-shell are $218.327$ and $202.436$ for USDB and SDNN interactions which shows the latter is approximately 1.1 factor better than the former one. For  $^{24}{Mg}$ isotope, the first excited $0^{+}$ state energies predicted are higher from the experimental one for both USDB and SDNN interactions. However, the situation is inverted for $^{28}{Si}$ isotope.

\begin{figure}
\centering
\includegraphics[height=23cm]{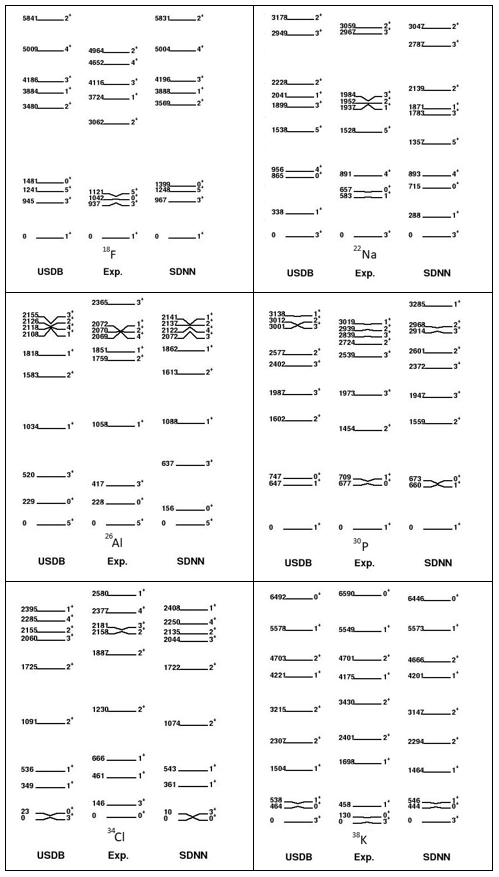}
\caption{Experimental and calculated energies for odd-odd $N=Z$ sd-shell nuclei by using USDB and SDNN interactions.}
\label{fig2}
\end{figure}

In the second part of this study, we have calculated the first ten positive parity energy levels and spins of odd-odd $N=Z$ nuclei in sd-shell by using both USDB and SDNN interaction. All the ground state spin/parity values were estimated correctly by means of our new SDNN interaction. Whereas USDB interaction could not reproduce the $J^{\pi}$ experimental ($0^{+}$), it gives $3^{+}$ for this state in $^{34}{Cl}$ isotope. As shown in Fig.\ref{fig2}, the calculated energy level deviation from the experimental values \cite{nudat} are close to each other for both interactions. The rms deviations for all $N=Z$ isotopes in sd-shell are $208.255$ and $205.363$ for USDB and SDNN interactions. 

\begin{figure}
\centering
\includegraphics[height=8cm]{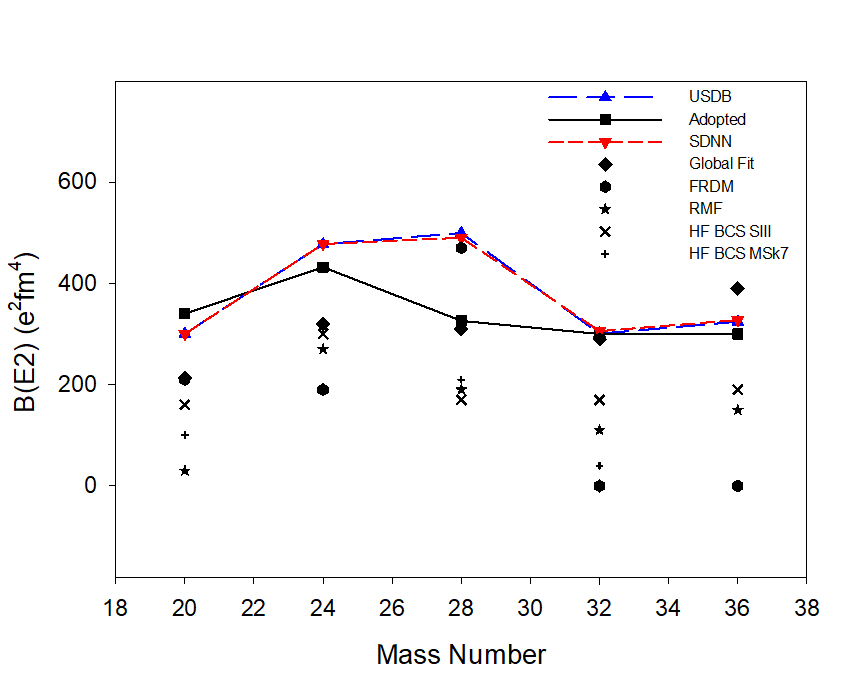}
\caption{Experimental and theoretical B(E2) values}
\label{fig3}
\end{figure}

We have also calculated $B(E2)\uparrow$ values for even-even $N=Z$ nuclei which are shown in Fig.\ref{fig3}. As it can be clearly seen in this figure, the obtained results are very close to each other. The larger difference from adopted values, is observed in  $^{28}{Si}$ isotope. In the literature, all the models, except FRDM, give smaller results for $B(E2)\uparrow$ than the adopted value for this isotope. But in our calculations, we have obtained a $B(E2)\uparrow$ value larger than the adopted one, by both interactions, where SDNN gives slightly closer results. For the other isotopes, the shell model results are closer to the adopted values than the results obtained from FRDM, RMF, ETFSI and HF+BCS models \cite{Raman}.

\begin{figure}
\centering
\includegraphics[height=6cm]{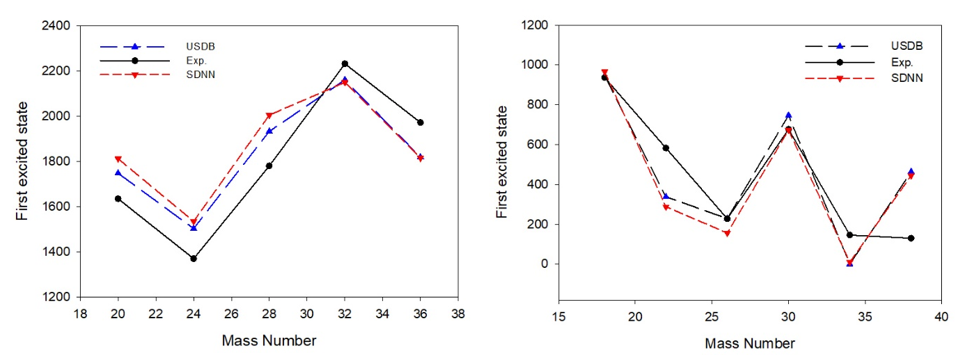}
\caption{Experimental and calculated first excited state energies for even-even (left) and odd-odd (right) isotopes}
\label{fig4}
\end{figure}

In Fig.\ref{fig4}, the first excited state energies are shown for even-even and odd-odd $N=Z$ sd nuclei. Calculated results are very close to each other and to experimental values for even-even isotopes. But for odd-odd nuclei, the deviations from the experimental values are very large for $^{22}{Na}$ and $^{38}{K}$ isotopes for USDB and SDNN interactions, while for the other isotopes the results are very close to experiment.

\newpage
\section{Conclusions}
In this study, a new set of two body matrix elements has been generated using USDB interaction. In this generation, the artificial neural network method has been used. The ground states spin and excited energies levels of $N=Z$ nuclei in sd shell have been calculated by means of the SDNN new interaction and the USDB original one. All the ground sate spins have been assigned correctly from SDNN according to the available literature values. The excited states of the nuclei which are calculated by the new interaction are closer to the experimental energies in comparison to the USDB original interaction. The rms deviations for $N=Z$ isotopes in sd-shell are $208.255$ and $205.363$ for USDB and SDNN interactions respectively. These results show that ANN method is capable to improve the TBMEs for the interactions used in nuclear structure studies.

\subsection{Acknowledgments}
This work is supported by the Scientific Research Project Fund of Sivas Cumhuriyet University under the project number F-616.

\end{document}